# The Join Levels of the Trotter-Weil Hierarchy are Decidable

Manfred Kufleitner*     Alexander Lauser*

University of Stuttgart, FMI
{kufleitner,lauser}@fmi.uni-stuttgart.de

**Abstract.** The variety **DA** of finite monoids has a huge number of different characterizations, ranging from two-variable first-order logic $\text{FO}^2$ to unambiguous polynomials. In order to study the structure of the subvarieties of **DA**, Trotter and Weil considered the intersection of varieties of finite monoids with bands, *i.e.*, with idempotent monoids. The varieties of idempotent monoids are very well understood and fully classified. Trotter and Weil showed that for every band variety **V** there exists a unique maximal variety **W** inside **DA** such that the intersection with bands yields the given band variety **V**. These maximal varieties **W** define the Trotter-Weil hierarchy. This hierarchy is infinite and it exhausts **DA**; induced by band varieties, it naturally has a zigzag shape.

Kufleitner and Weil have shown that it is possible to ascend the corners of the Trotter-Weil hierarchy in terms of Mal'cev products with definite and reverse definite semigroups and also in terms of so-called rankers. Using a result of Straubing, one can climb up the intersection levels of the Trotter-Weil hierarchy in terms of weakly iterated block products with $\mathcal{J}$-trivial monoids. In their paper, Trotter and Weil have shown that the corners and the intersection levels of this hierarchy are decidable.

In this paper, we give a single identity of omega-terms for every join level of the Trotter-Weil hierarchy; this yields decidability. Moreover, we show that the join levels and the subsequent intersection levels do not coincide. Almeida and Azevedo have shown that the join of $\mathcal{R}$-trivial and $\mathcal{L}$-trivial finite monoids is decidable; this is the first non-trivial join level of the Trotter-Weil hierarchy. We extend this result to the other join levels of the Trotter-Weil hierarchy. At the end of the paper, we give two applications. First, we show that the hierarchy of deterministic and codeterministic products is decidable. And second, we show that the direction alternation depth of unambiguous interval logic is decidable.

*The authors acknowledge the support by the German Research Foundation (DFG) under grant DI 435/5-1.

# 1 Introduction

The lattice of band varieties was classified independently by Birjukov, Fennemore, and Gerhard [3, 6, 7]. For the purpose of this paper, a band is a finite idempotent monoid; and a variety is a class of finite monoids which is closed under submonoids, homomorphic images, and finite direct products. We denote the variety of all bands by **B**. The relation between the band varieties can be found on the left-hand side of Figure 1 where we use the notation of [8, 21]. A famous supervariety of **B** is **DA**, the class of all finite monoids such that every regular $\mathcal{D}$-class is an aperiodic semigroup. This variety appears at a huge number of different occasions, see *e.g.* [19, 4] for overviews. The most prominent results along this line of work are the following: A language is definable in two-variable first-order logic $\text{FO}^2$ if and only if its syntactic monoid is in **DA** [20] if and only if it is a disjoint union of unambiguous monomials [16].

Trotter and Weil studied the lattice of all subvarieties of **DA**. This lattice is uncountably infinite whereas the lattice of band varieties is countably infinite. Considering the bands inside the subvarieties of **DA** led to the following result. For every given band variety **V** they showed that there exists a unique maximal variety $\mathbf{W} \subseteq \mathbf{DA}$ such that $\mathbf{V} = \mathbf{W} \cap \mathbf{B}$, *cf.* [21]. These maximal varieties **W** constitute the Trotter-Weil hierarchy. Its structure is depicted on the right-hand side of Figure 1. The zigzag shape gives rise to the following notions. We say the varieties $\mathbf{R}_m$ and $\mathbf{L}_m$ are the *corners*, varieties of form $\mathbf{R}_m \cap \mathbf{L}_m$ are the *intersection levels*, and $\mathbf{R}_m \vee \mathbf{L}_m$ are the *join levels* of the Trotter-Weil hierarchy. Later, Kufleitner and Weil showed that there exist several different ways of climbing up along the corners of the Trotter-Weil hierarchy. One possibility is in terms of Mal'cev products with definite and reverse definite semigroups [10]; and another possibility uses condensed rankers [12]. The concept of condensed rankers is a refinement of the rankers of Weis and Immerman [22] and the turtle programs of Schwentick, Thérien, and Vollmer [17]. Condensed rankers are very similarly to the unambiguous interval logic of Lodaya, Pandya, and Shah [13], which in turn gives yet another way of climbing up the Trotter-Weil hierarchy.

Kufleitner and Weil showed that the $\text{FO}^2$ quantifier alternation hierarchy and the Trotter-Weil hierarchy are interwoven [12]. Only recently, they tightened this connection: A language is definable in $\text{FO}^2$ with $m$ blocks of quantifiers if and only if it is recognizable by a monoid in $\mathbf{R}_{m+1} \cap \mathbf{L}_{m+1}$, *cf.* [11]. Therefore, a result of Straubing on the $\text{FO}^2$ alternation hierarchy shows that it is possible to climb up the Trotter-Weil hierarchy along the intersection levels by using weakly iterated block products of $\mathcal{J}$-trivial monoids [18].

Pin showed that the algebraic operations of taking Mal'cev products with definite and reverse definite semigroups admit language counterparts by means of deterministic and codeterministic products [14], see also [15, 16]. Thus a language over the alphabet $A$ is recognizable by a monoid in **DA** if and only if it is in the closure of languages $B^*$ for $B \subseteq A$ under Boolean operations and deterministic and codeterministic products [12]. This naturally defines a hierarchy of languages inside **DA**. Let $\mathcal{W}_1$ be the Boolean closure of languages of the form $B^*$, and let $\mathcal{W}_{m+1}$ be the Boolean closure of deterministic and of codeterministic products of languages in $\mathcal{W}_m$. Now, a language is in $\mathcal{W}_m$ if and only if it is recognizable by a monoid in $\mathbf{R}_m \vee \mathbf{L}_m$. In particular, this is an infinite hierarchy which exhausts the class of all languages **DA**-recognizable languages. Note that if one would replace deterministic and codeterministic products by unambiguous products, then Schützenberger's result [16] shows that the resulting hierarchy collapses at level 2.

Our main result is a single identity of omega-terms for each of the varieties $\mathbf{R}_m \vee \mathbf{L}_m$. It follows that membership in $\mathbf{R}_m \vee \mathbf{L}_m$ is decidable. Since $\mathbf{R}_2$ is the class of $\mathcal{R}$-trivial monoids and $\mathbf{L}_2$ is the class of $\mathcal{L}$-trivial monoids, this extends the decidability result for the join of $\mathcal{R}$-trivial and $\mathcal{L}$-trivial monoids by Almeida and Azevedo [2] to the other join levels of the Trotter-Weil hierarchy. In fact, the Almeida-Azevedo result is the base of our proof. As a byproduct, we give a new single identity of omega-terms for the corners of the Trotter-Weil hierarchy. Different identities were obtained by Trotter and Weil [21]. We complement our main result by showing that, for every $m \geq 2$, the variety $\mathbf{R}_m \vee \mathbf{L}_m$ is strictly contained in $\mathbf{R}_{m+1} \cap \mathbf{L}_{m+1}$. Note that for band varieties, the join levels coincide with the subsequent intersection levels, see *e.g.* [8].

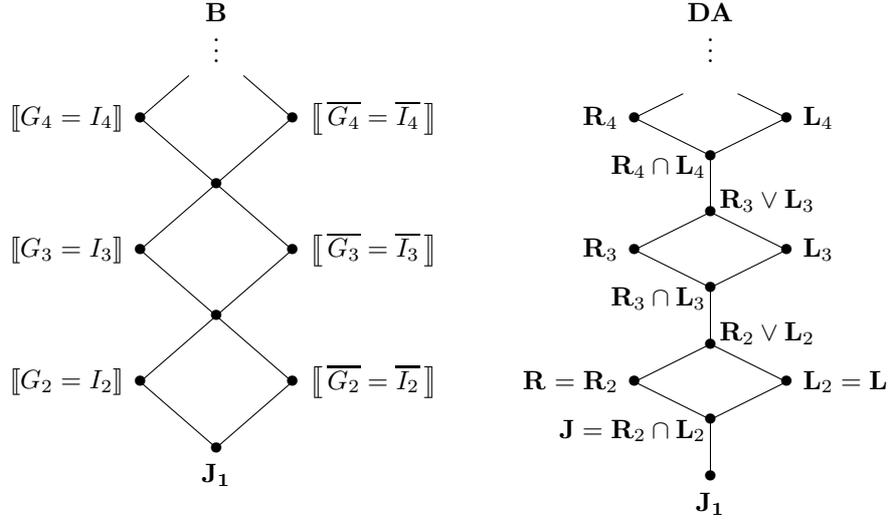

Figure 1: The band hierarchy and the Trotter-Weil hierarchy

We give two applications. The first one is decidability of the hierarchy $\mathcal{W}_m$ of deterministic and codeterministic products. This easily follows from our main result (Theorem 1) and from Pin's characterization of deterministic and codeterministic products [14]. And the second application is the following: For every integer $m$ it is decidable whether a given regular language $L$ is definable in unambiguous interval logic with at most $m$ direction alternations (see Section 7.2 for definitions).

## 2 Preliminaries

**Words and Languages.** Throughout this paper we let $A$ be a finite alphabet. The set of finite words over $A$ is denoted by $A^*$. It is the free monoid over $A$. The *empty word* 1 is the neutral element. As usual, we set $A^+ = A^* \setminus \{1\}$. The *length* of a word $u = a_1 \cdots a_n$ with $a_i \in A$ is $|u| = n$, and its *alphabet* (also known as its *content*) is the set $\alpha(u) = \{a_1, \ldots, a_n\}$. A homomorphism $\varphi : A^* \to M$ to a monoid $M$ *recognizes* a language $L \subseteq A^*$ if $\varphi^{-1}\varphi(L) = L$. A monoid $M$ *recognizes* $L \subseteq A^*$ if there exists a homomorphism $\varphi : A^* \to M$ which recognizes $L$. Every language admits a unique minimal monoid $M$ which recognizes $L$. This monoid is called the *syntactic monoid* of $L$, see *e.g.* [15] for details. A language $L$ is *regular* (or *recognizable*) if it is recognized by a finite monoid.

**Finite Monoids.** Let $M$ be a monoid. An element $e \in M$ is *idempotent* if $e^2 = e$. If $M$ is finite, then there exists a positive integer $\omega \in \mathbb{N}$ such that $x^\omega$ is idempotent for all $x \in M$. The idempotent $x^\omega$ generated by $x$ is unique. *Green's relations* $\mathcal{R}$ and $\mathcal{L}$ are an important tool for describing the structure of monoids. For $x, y \in M$ we define

$$x \mathrel{\mathcal{R}} y \text{ if and only if } xM = yM, \quad x \leq_{\mathcal{R}} y \text{ if and only if } xM \subseteq yM,$$
$$x \mathrel{\mathcal{L}} y \text{ if and only if } Mx = My, \quad x \leq_{\mathcal{L}} y \text{ if and only if } Mx \subseteq My.$$

We frequently use these relations as follows: The relation $x \leq_{\mathcal{R}} y$ holds if and only if there exists $z \in M$ such that $x = yz$, Similarly, $x \leq_{\mathcal{L}} y$ if and only if there exists $z \in M$ such that $x = zy$. We say that a monoid $M$ is $\mathcal{R}$-*trivial* (resp. $\mathcal{L}$-*trivial*) if $\mathcal{R}$ (resp. $\mathcal{L}$) is the identity relation on $M$.

Every homomorphism $\varphi : A^* \to M$ induces a congruence $\equiv_\varphi$ on $A^*$ by setting $x \equiv_\varphi y$ if $\varphi(x) = \varphi(y)$. Now, the submonoid $\varphi(A^*)$ of $M$ is isomorphic to $A^*/\equiv_\varphi$. For defining the Trotter-Weil hierarchy, we introduce the congruences $\sim_K$ and $\sim_D$ on $M$. We let $x \sim_K y$ if for all idempotents $e$ of $M$ the following implication holds:

$$\text{if } ex \mathrel{\mathcal{R}} e \text{ or } ey \mathrel{\mathcal{R}} e, \text{ then } ex = ey.$$

Using more semigroup theoretic notions, the meaning $x \sim_K y$ is that, for every regular $\mathcal{D}$-class $D$, the right translations by $x$ and by $y$ define the same partial functions on $D$, see e.g. [9]. The left-right dual $\sim_D$ is defined by $x \sim_D y$ if for all idempotents $e$ of $M$ we have that if $xe \mathrel{\mathcal{L}} e$ or $ye \mathrel{\mathcal{L}} e$, then $xe = ye$.

**Varieties of Finite Monoids.** A *variety* is a class of finite monoids which is closed under submonoids, homomorphic images, and finite direct products. The empty direct product of monoids yields the one-elements monoid $\{1\}$. Thus the monoid $\{1\}$ is contained in every variety. The *join* $\mathbf{V} \vee \mathbf{W}$ of two varieties $\mathbf{V}$ and $\mathbf{W}$ is the smallest variety containing both $\mathbf{V}$ and $\mathbf{W}$. A *language variety* is a class of regular languages which is closed under Boolean operations, inverse homomorphic images, and residuals. More formally, the languages in a language variety are parametrized by the alphabet, but in order to keep the notations in this paper simple, we use this distinction only implicitly. Eilenberg has shown that there is a one-to-one correspondence between language varieties and varieties of finite monoids [5]. This correspondence is defined by the following mutually inverse relationships: To every language variety $\mathcal{V}$ one can assign the variety of finite monoids generated by the syntactic monoids of languages in $\mathcal{V}$; and to every variety $\mathbf{V}$ of finite monoids one can assign the languages recognized by the monoids in $\mathbf{V}$.

Identities of omega-terms are a very common way of defining varieties. We inductively define *omega-terms* over a set of variables $\Sigma$. The empty word 1 and every $x \in \Sigma$ is an omega-term; and if $u$ and $v$ are omega-terms, then so are $uv$ and $(u)^\omega$. Every homomorphism $\varphi : \Sigma^* \to M$ to a finite monoid $M$ naturally extends to omega-terms by setting $\varphi(u^\omega) = \varphi(u)^\omega$, i.e., $\varphi(u^\omega)$ is the idempotent generated by $\varphi(u)$. Let $u$ and $v$ be two omega-terms over $\Sigma$. A finite monoid $M$ *satisfies* the identity $u = v$ if for every homomorphism $\varphi : \Sigma^* \to M$ we have $\varphi(u) = \varphi(v)$. The class of all finite monoids which satisfy $u = v$ is denoted by $[\![u = v]\!]$. For all omega-terms $u$ and $v$, the class $[\![u = v]\!]$ is a variety. We will use the following varieties

in this paper:

$$\begin{aligned}
\mathbf{DA} &= [\![(xy)^\omega x(xy)^\omega = (xy)^\omega]\!] \\
\mathbf{R} &= [\![(zx)^\omega z = (zx)^\omega]\!] \\
\mathbf{L} &= [\![z(yz)^\omega = (yz)^\omega]\!] \\
\mathbf{J} &= \mathbf{R} \cap \mathbf{L} \\
\mathbf{B} &= [\![x^2 = x]\!] \\
\mathbf{J_1} &= [\![xy = yx]\!] \cap \mathbf{B} \\
\mathbf{W}_m &= [\![e_m \cdots e_1 z f_1 \cdots f_m = e_m \cdots e_1 f_1 \cdots f_m]\!] \quad \text{where } e_1 = f_1 = 1 \text{ and} \\
& \quad e_{i+1} = (e_i \cdots e_1 z f_1 \cdots f_i x_i)^\omega \text{ and } f_{i+1} = (y_i e_i \cdots e_1 z f_1 \cdots f_i)^\omega.
\end{aligned}$$

In particular, we have $\mathbf{W}_2 = [\![(zx_1)^\omega z(y_1 z)^\omega = (zx_1)^\omega (y_1 z)^\omega]\!]$. A monoid is in $\mathbf{R}$ if and only if it is $\mathcal{R}$-trivial, and it is in $\mathbf{L}$ if and only if it is $\mathcal{L}$-trivial, see *e.g.* [15]. The elements of $\mathbf{J}$ are called $\mathcal{J}$-*trivial* monoids.

If $\mathbf{V}$ is a variety, then $\mathbf{K}\textcircled{m}\mathbf{V}$ contains all monoids $M$ such that $M/\!\sim_K$ is in $\mathbf{V}$. Symmetrically, $M$ is in $\mathbf{D}\textcircled{m}\mathbf{V}$ if $M/\!\sim_D$ is in $\mathbf{V}$. Usually, the Mal'cev products $\mathbf{K}\textcircled{m}\mathbf{V}$ and $\mathbf{D}\textcircled{m}\mathbf{V}$ are defined using relational morphisms, but the definition given here is equivalent [9]. We are now ready to define the *Trotter-Weil hierarchy*. We set $\mathbf{R}_2 = \mathbf{R}$ and $\mathbf{L}_2 = \mathbf{L}$, and for $m \geq 2$ we let $\mathbf{R}_{m+1} = \mathbf{K}\textcircled{m}\mathbf{L}_m$ and $\mathbf{L}_{m+1} = \mathbf{D}\textcircled{m}\mathbf{R}_m$. There are several possible extensions to the first level so as to obtain the same hierarchy for the higher levels: we have $\mathbf{K}\textcircled{m}\mathbf{V} = \mathbf{R}_2$ and $\mathbf{D}\textcircled{m}\mathbf{V} = \mathbf{L}_2$ for every variety $\mathbf{V}$ with $\mathbf{J_1} \subseteq \mathbf{V} \subseteq \mathbf{J}$, see *e.g.* [15]. It depends on the context what the most natural choice is, and we therefore start the hierarchy at level 2. The structure of the Trotter-Weil hierarchy is depicted on the right-hand side of Figure 1. It is well-known that $\bigcup_m \mathbf{R}_m = \bigcup_m \mathbf{L}_m = \mathbf{DA}$, see [10]. The main purpose of this paper is to show

$$\mathbf{W}_m = \mathbf{R}_m \vee \mathbf{L}_m \subsetneq \mathbf{R}_{m+1} \cap \mathbf{L}_{m+1}.$$

The containment $\mathbf{R}_m \vee \mathbf{L}_m \subseteq \mathbf{R}_{m+1} \cap \mathbf{L}_{m+1}$ is straightforward, see [12, Corollary 3.19]. Almeida and Azevedo [2] have shown that $\mathbf{R} \vee \mathbf{L} = \mathbf{W}_2$, see also [1].

## 3 Identities for the Corners

The following lemma is one of the main properties of monoids in $\mathbf{DA}$. It basically says that, inside $\mathbf{DA}$, whether or not $u \mathcal{R} ua$ only depends on $a$ and the $\mathcal{R}$-class of $u$, but not on the element $u$. Symmetrically, for monoids in $\mathbf{DA}$, whether or not $u \mathcal{L} au$ holds only depends on $a$ and the $\mathcal{L}$-class of $u$.

**Lemma 1** *Let $M \in \mathbf{DA}$ and let $u, v, a \in M$. If $v \mathcal{R} u \mathcal{R} ua$, then $v \mathcal{R} va$. If $v \mathcal{L} u \mathcal{L} au$, then $v \mathcal{L} av$.*

*Proof:* The second implication is left-right symmetric to the first as $(xy)^\omega x(xy)^\omega = (xy)^\omega$ if and only if $(xy)^\omega x(xy)^\omega x = (xy)^\omega x$ if and only if $x(yx)^\omega x(yx)^\omega = x(yx)^\omega$ if and only if $(yx)^\omega x(yx)^\omega = (yx)^\omega$. Hence, it suffices to show the first part of the statement. Let $u \mathcal{R} v$ and $u \mathcal{R} ua$ and suppose $u = vx$, $v = uy$ and $u = uaz$. Then $v = v(xazy)^\omega = vx(azyx)^\omega azy(xazy)^{\omega-1}$. Since $M \in \mathbf{DA}$ we see $(azyx)^\omega = (azyx)^\omega azy(azyx)^\omega$ and therefore $v \in vx(azyx)^\omega azyaM = vaM$. Thus $v \mathcal{R} va$. □

The next lemma shows that we can apply Lemma 1 for $M \in \mathbf{W}_m$.

**Lemma 2** *For all $m \geq 2$ we have $\mathbf{W}_m \subseteq \mathbf{DA}$.*

*Proof:* Let $M \in \mathbf{W}_m$. Setting $x_i = y_i = z$ for all $i$ yields $e_i = f_i = z^\omega$ and consequently $z^\omega z = z^\omega$, that is, $M$ is aperiodic. With $x_i = y_i = y$ for all $i$ we get $e_i = (zy)^\omega$ and $f_i = (yz)^\omega$. The defining identity for $\mathbf{W}_m$ implies $(zy)^\omega(yz)^\omega = (zy)^\omega z$. Hence, $(yz)^\omega y(yz)^\omega = y(zy)^\omega(yz)^\omega = y(zy)^\omega z = (yz)^\omega$. The last equality relies on aperiodicity of $M$. □

The next proposition gives a new equational description of the corners of the Trotter-Weil hierarchy. This description immediately yields $\mathbf{R}_m \vee \mathbf{L}_m \subseteq \mathbf{W}_m$.

**Proposition 1** *Let $e_1 = f_1 = 1$ and for $i \geq 1$ let $e_{i+1} = (e_i \cdots e_1 z f_1 \cdots f_i x_i)^\omega$ and $f_{i+1} = (y_i e_i \cdots e_1 z f_1 \cdots f_i)^\omega$. Then for all $m \geq 2$ we have*

$$\mathbf{R}_m = [\![e_m \cdots e_1 z f_1 \cdots f_{m-1} = e_m \cdots e_1 f_1 \cdots f_{m-1}]\!],$$
$$\mathbf{L}_m = [\![e_{m-1} \cdots e_1 z f_1 \cdots f_m = e_{m-1} \cdots e_1 f_1 \cdots f_m]\!].$$

*Proof:* For $m = 2$ the claim is true by definition of $\mathbf{R}_2$ and $\mathbf{L}_2$. Let now $m \geq 3$. By left-right symmetry it suffices to show the statement for $\mathbf{R}_m$. First, we consider the inclusion from left to right. Let $M \in \mathbf{R}_m$. Then $M/\sim_K \in \mathbf{L}_{m-1}$ and by induction, $M/\sim_K$ satisfies $e_{m-2} \cdots e_1 z f_1 \cdots f_{m-1} = e_{m-2} \cdots e_1 f_1 \cdots f_{m-1}$, i.e., the elements $u = e_{m-2} \cdots e_1 z f_1 \cdots f_{m-1}$ and $v = e_{m-2} \cdots e_1 f_1 \cdots f_{m-1}$ satisfy $u \sim_K v$. Thus $e_{m-1} u \sim_K e_{m-1} v$. We have $e_m \mathcal{R} e_m e_{m-1} u$ because $e_m \leq_\mathcal{R} e_{m-1} u$. Hence, $e_m e_{m-1} u = e_m e_{m-1} v$ by definition of $\sim_K$.

Next, we show the inclusion from right to left. Let $M$ satisfy $e_m \cdots e_1 z f_1 \cdots f_{m-1} = e_m \cdots e_1 f_1 \cdots f_{m-1}$. We have $M \in \mathbf{DA}$ by Lemma 2. For $u = e_{m-2} \cdots e_1 z f_1 \cdots f_{m-1}$ and $v = e_{m-2} \cdots e_1 f_1 \cdots f_{m-1}$ we claim $u \sim_K v$. Then $M/\sim_K \in \mathbf{L}_{m-1}$ by induction which yields $M \in \mathbf{R}_m$. Consider an idempotent element $e \in M$ such that $eu \mathcal{R} e$. Note that by Lemma 1, we have $e \mathcal{R} eu$ if and only if $e \mathcal{R} ev$. There exists $x_{m-1} \in M$ with $e = eux_{m-1}$. Let $x_{m-2} = f_{m-1} x_{m-1}$, let $e_{m-1} = (e_{m-2} \cdots e_1 z f_1 \cdots f_{m-2} x_{m-2})^\omega$, and let $e_m = (e_{m-1} e_{m-2} \cdots e_1 z f_1 \cdots f_{m-1} x_{m-1})^\omega$. We have $e_m = e_{m-1} = (u x_{m-1})^\omega$. By choice of $M$ we have $e_m e_{m-1} u = e_m e_{m-1} v$ and hence, $eu = e e_m e_{m-1} u = e e_m e_{m-1} v = ev$. This shows $u \sim_K v$. □

## 4 Rankers

Condensed rankers are important for the proofs of both of our main results Theorem 1 and Theorem 2. A *ranker* is a nonempty word over the alphabet $Z_A = \{\mathsf{X}_a, \mathsf{Y}_a \mid a \in A\}$ with $2\,|A|$ letters. The set $Z_A$ is partitioned into $Z_A = X_A \cup Y_A$ with $X_A = \{\mathsf{X}_a \mid a \in A\}$ and $Y_A = \{\mathsf{X}_a \mid a \in A\}$. Every ranker is interpreted as a sequence of instructions of the form "go to the next $a$-position" and "go to the previous $a$-position". More formally, for $u = a_1 \cdots a_n \in A^*$ and $x \in \{0, \ldots, n+1\}$ we let

$$\mathsf{X}_a(u, x) = \min\{y \mid y > x \text{ and } a_y = a\}, \qquad \mathsf{X}_a(u) = \mathsf{X}_a(u, 0),$$
$$\mathsf{Y}_a(u, x) = \max\{y \mid y < x \text{ and } a_y = a\}, \qquad \mathsf{Y}_a(u) = \mathsf{Y}_a(u, n+1).$$

Here, both the minimum and the maximum of the empty set are undefined. The modality $\mathsf{X}_a$ is for "neXt-$a$" and $\mathsf{Y}_a$ is for "Yesterday-$a$". For a ranker $r = \mathsf{Z}s$ with $\mathsf{Z} \in Z_A$ we set $r(u) = s(u, \mathsf{Z}(u))$ and $r(u,x) = s(u, \mathsf{Z}(u,x))$. In particular, the instructions of a ranker are executed from left to right. Every ranker $r$ either defines a unique position in a word $u$, or it is undefined on $u$. For example, $\mathsf{X}_a\mathsf{Y}_b\mathsf{X}_c(bca) = 2$ and $\mathsf{X}_a\mathsf{Y}_b\mathsf{X}_c(bac) = 3$ whereas $\mathsf{X}_a\mathsf{Y}_b\mathsf{X}_c(abac)$ and $\mathsf{X}_a\mathsf{Y}_b\mathsf{X}_c(bcba)$ are undefined. A ranker $r$ is *condensed* on $u$ if it is defined and, during the execution of $r$, no previously visited position is overrun [12]. More formally, let $r = \mathsf{Z}_1 \cdots \mathsf{Z}_k$ with $\mathsf{Z}_i \in Z_A$ be defined on $u$ and let $x_i = \mathsf{Z}_1 \cdots \mathsf{Z}_i(u)$ be the position reached after $i$ instructions. Then $r$ is *condensed* on $u$ if for every $i \leq k-1$ we have that either all positions $x_{i+1}, \ldots, x_k$ are greater than $x_i$ or all positions $x_{i+1}, \ldots, x_k$ are smaller than $x_i$. By definition, every ranker which is condensed on $u$ is also defined on $u$, but the converse is not true. For example, $\mathsf{X}_a\mathsf{Y}_b\mathsf{X}_c$ is condensed on *bca* but not on *bac*. Let

$$L_c(r) = \{u \in A^* \mid r \text{ is condensed on } u\}.$$

The *depth* of a ranker is its length as a word. A *block* of a ranker is a maximal factor either in $X_A^+$ or $Y_A^+$. A ranker with $m$ blocks changes direction $m-1$ times. By $R_{m,n}$ we denote the rankers with depth at most $n$ and with up to $m$ blocks. Let $R_m = \bigcup_n R_{m,n}$. We set $R_{m,n}^\mathsf{X} = (R_{m,n} \cap X_A Z_A^*) \cup R_{m-1,n-1}$ and $R_{m,n}^\mathsf{Y} = (R_{m,n} \cap Y_A Z_A^*) \cup R_{m-1,n-1}$. We write $u \rhd_{m,n} v$ (resp. $u \lhd_{m,n} v$) if $u$ and $v$ are condensed on the same rankers in $R_{m,n}^\mathsf{X}$ (resp. $R_{m,n}^\mathsf{Y}$). Let $u \equiv_{m,n} v$ if $u$ and $v$ are condensed on the same rankers from $R_{m,n}$, i.e., if both $u \rhd_{m,n} v$ and $u \lhd_{m,n} v$.

The following result of Kufleitner and Weil shows that condensed rankers can be used for defining the languages corresponding to the Trotter-Weil hierarchy [12].

**Proposition 2** *For every $n \in \mathbb{N}$ we have $A^*/\rhd_{m,n} \in \mathbf{R}_m$ and $A^*/\lhd_{m,n} \in \mathbf{L}_m$. For every homomorphism $\varphi : A^* \to M$ with $M \in \mathbf{R}_m$ (resp. $M \in \mathbf{L}_m$) there exists $n \in \mathbb{N}$ such that $u \rhd_{m,n} v$ (resp. $u \lhd_{m,n} v$) implies $\varphi(u) = \varphi(v)$ for all $u, v \in A^*$.*

*Proof:* See [12, Theorem 3.21]. □

This leads to the following corollary, which is the main motivation for considering condensed rankers in this paper.

**Corollary 1** *For every $n \in \mathbb{N}$ we have $A^*/\equiv_{m,n} \in \mathbf{R}_m \vee \mathbf{L}_m$. For every homomorphism $\varphi : A^* \to M$ with $M \in \mathbf{R}_m \vee \mathbf{L}_m$ there exists $n \in \mathbb{N}$ such that $u \equiv_{m,n} v$ implies $\varphi(u) = \varphi(v)$ for all $u, v \in A^*$.* □

**Lemma 3** *Let $r \in X_A^+$ and $s \in Y_A^+$. If $u \equiv_{1,|r|+|s|} v$, then $r(u) < s(u)$ if and only if $r(v) < s(v)$ and $r(u) > s(u)$ if and only if $r(v) > s(v)$.*

*Proof:* Note that rankers in $X_A^+ \cup Y_A^+$ are condensed on $u$ if and only if they are defined on $u$. Let $s = \mathsf{Y}_{a_1} \cdots \mathsf{Y}_{a_k}$. We have $r(u) < s(u)$ if and only if $r\mathsf{X}_{a_k} \cdots \mathsf{X}_{a_1}(u)$ is defined if and only if $r\mathsf{X}_{a_k} \cdots \mathsf{X}_{a_1}(v)$ is defined (by assumption) if and only if $r(v) < s(v)$. Assume now that $r(u) > s(u)$ and $r(v) \leq s(v)$. If $r(v) < s(v)$, then the above implies $r(u) < s(u)$, a contradiction. It remains the case $r(v) = s(v)$. Then $r\mathsf{X}_{a_2} \cdots \mathsf{X}_{a_k}$ is defined on $v$ but not on $u$, which is a contradiction. □

**Lemma 4** *Let $m \geq 2$, $u = u_0 a u_1$, and $v = v_0 a v_1$.*

1. *If $a \notin \alpha(u_0) \cup \alpha(v_0)$ and $u \rhd_{m,n} v$, then $u_0 \rhd_{m,n-1} v_0$ and $u_1 \rhd_{m,n-1} v_1$.*
2. *If $a \notin \alpha(u_0) \cup \alpha(v_0)$ and $u \lhd_{m,n} v$, then $u_1 \lhd_{m,n-1} v_1$.*
3. *If $a \notin \alpha(u_1) \cup \alpha(v_1)$ and $u \lhd_{m,n} v$, then $u_0 \lhd_{m,n-1} v_0$ and $u_1 \lhd_{m,n-1} v_1$.*
4. *If $a \notin \alpha(u_1) \cup \alpha(v_1)$ and $u \rhd_{m,n} v$, then $u_0 \rhd_{m,n-1} v_0$.*

*Proof:* See [12, Proposition 3.6 and Lemma 3.7]. □

The following lemma relativizes condensed rankers to factors between ranker positions. It will be the main combinatorial ingredient for the inductive step in showing that $\mathbf{W}_m$ is contained in the join of $\mathbf{R}_m$ and $\mathbf{L}_m$.

**Lemma 5** *Let $r \in X_A^*$ and $s \in Y_A^*$, and let $u = u_0 a_1 u_1 \cdots a_\ell u_\ell$ and $v = v_0 a_1 v_1 \cdots a_\ell v_\ell$ with $a_i \in A$ such that the $a_i$'s correspond to the positions visited by $r$ and $s$, i.e.,*

1. *if $t$ is a nonempty prefix of either $r$ or $s$, then there exists $i \in \{1, \ldots, \ell\}$ such that $t(u) = |u_0 a_1 \cdots u_{i-1} a_i|$ and $t(v) = |v_0 a_1 \cdots v_{i-1} a_i|$, and*
2. *for all $i \in \{1, \ldots, \ell\}$ there exists a nonempty prefix $t$ of $r$ or $s$ such that $t(u) = |u_0 a_1 \cdots u_{i-1} a_i|$ and $t(v) = |v_0 a_1 \cdots v_{i-1} a_i|$.*

*If $u \equiv_{m,n+\ell+1} v$ for $m \geq 3$, then $u_i \equiv_{m-1,n} v_i$ for all $i \in \{0, \ldots, \ell\}$.*

*Proof:* Throughout the proof, we silently rely on Lemma 4. The proof is by induction on $\ell$. If $\ell = 0$, then the claim is trivially true. Let now $\ell > 0$. By symmetry, we can assume that $r$ is nonempty. We distinguish whether or not the position of $a_1$ is visited by $r$. First, suppose $r = \mathsf{X}_{a_1} r'$ for $r' \in X_A^*$. We have $u_0 \triangleright_{m,n+\ell} v_0$, and hence $u_0 \equiv_{m-1,n} v_0$. The words $u' = u_1 a_2 \cdots u_\ell$ and $v' = v_1 a_2 \cdots v_\ell$ satisfy $u' \triangleright_{m,n+\ell} v'$ and $u' \triangleleft_{m,n+\ell} v'$. Thus $u' \equiv_{m,n+\ell} v'$. Without loss of generality we can assume $s(u) \neq |u_0 a_1|$, i.e., the position of $a_1$ is not visited by both $r$ and $s$; otherwise we delete the last modality of $s$. Induction hypothesis for $u'$ and $v'$ with the rankers $r'$ and $s$ yields $u_i \equiv_{m-1,n} v_i$ for all $i \in \{1, \ldots, \ell\}$.

The remaining case is that $r$ does not start with an $\mathsf{X}_{a_1}$-modality. Let $r = \mathsf{X}_a r'$ for $r' \in X_A^*$. There exists $k$ such that $a = a_k$ and $a \notin \alpha(u_0 a_1 \cdots u_{k-1}) \cup \alpha(v_0 a_1 \cdots v_{k-1})$. Moreover, we can write $s = s' \mathsf{Y}_{a_{k-1}} \cdots \mathsf{Y}_{a_1}$ for some $s' \in Y_A^*$. Let $u' = u_k a_{k+1} \cdots u_\ell$ and $v' = v_k a_{k+1} \cdots v_\ell$. We have $u' \equiv_{m,n+\ell} v'$. Hence $r'$ and $s'$ satisfy the hypothesis for $u'$ and $v'$ (as before, without loss of generality we can assume $s'(u) \neq |u_0 a_1 \cdots u_{k-1} a_k|$). By induction we have $u_i \equiv_{m-1,n} v_i$ for all $i \geq k$. Let $u'' = u_0 a_1 \cdots u_{k-2}$ and $v'' = v_0 a_1 \cdots v_{k-2}$. We have $u'' a_{k-1} u_{k-1} \triangleright_{m,n+\ell} v'' a_{k-1} v_{k-1}$ and thus $u'' \triangleright_{m,n+k-1} v''$. An $(|s'|+1)$-fold application of Lemma 4 to $u$ and $v$ leads to $u'' \triangleleft_{m,n+k-1} v''$. Therefore $u'' \equiv_{m,n+k-1} v''$ and by induction (using the empty ranker and $\mathsf{Y}_{a_{k-2}} \cdots \mathsf{Y}_{a_1}$) we see that $u_i \equiv_{m-1,n} v_i$ for all $i \leq k-2$.

It remains to show $u_{k-1} \equiv_{m-1,n} v_{k-1}$. Since $u'' a_{k-1} u_{k-1} \triangleright_{m,n+\ell} v'' a_{k-1} v_{k-1}$ we have $u'' a_{k-1} u_{k-1} \triangleleft_{m-1,n+\ell-1} v'' a_{k-1} v_{k-1}$ and, since $m \geq 3$, we obtain $u_{k-1} \triangleleft_{m-1,n} v_{k-1}$. Consider $i$ such that $s'(u) = |u_0 a_1 \cdots u_i a_{i+1}|$. Then we have $i \geq k-1$ and $u_0 a_1 \cdots u_i \triangleleft_{m,n+i+1} v_0 a_1 \cdots v_i$. Now, $u_{k-1} a_k \cdots u_i \triangleleft_{m,n+i} v_{k-1} a_k \cdots v_i$ since $a_{k-1} \notin \alpha(u_{k-1} a_k \cdots u_i) \cup \alpha(v_{k-1} a_k \cdots v_i)$. It follows $u_{k-1} a_k \cdots u_i \triangleright_{m-1,n+i-1} v_{k-1} a_k \cdots v_i$ and, since $a_k \notin \alpha(u_{k-1}) \cup \alpha(v_{k-1})$ and $m \geq 3$, we conclude $u_{k-1} \triangleright_{m-1,n} v_{k-1}$. This shows $u_{k-1} \equiv_{m-1,n} v_{k-1}$. □

## 5 Identities for the Join Levels

This section contains our main contribution Theorem 1. The proof of the inclusion $\mathbf{W}_m \subseteq \mathbf{R}_m \vee \mathbf{L}_m$ is by induction on $m$. The base case $m = 2$ was shown by Almeida and Azevedo [2]. The induction step connects condensed rankers with the variety $\mathbf{W}_m$. Essentially, Lemma 6 gives the first half of the induction step. The second half, which is Lemma 8, relies on Lemma 7 and the combinatorial properties of condensed rankers in Lemma 5. Finally, Corollary 1 yields the connection between condensed rankers and the join levels of the Trotter-Weil hierarchy.

**Lemma 6** *Let $m \geq 3$ and let $\varphi : A^* \to M$ be a homomorphism with $M \in \mathbf{W}_m$. Suppose that for every homomorphism $\mu : A^* \to N$ with $N \in \mathbf{W}_{m-1}$ there exists $n' \in \mathbb{N}$ such that $u \equiv_{m-1,n'} v$ implies $u \equiv_\mu v$ for all $u, v \in A^*$. Then there exists $n \in \mathbb{N}$ such that for all $u, v, x, y \in A^*$ the following implication holds:*

$$u \equiv_{m-1,n} v, \ \varphi(x) \mathcal{R} \varphi(xu), \ \varphi(y) \mathcal{L} \varphi(vy) \ \Rightarrow \ xuy \equiv_\varphi xvy.$$

*Proof:* To shorten notation, we extend relations $\mathcal{G}$ on $M$ to words by setting $u \mathcal{G} v$ if and only if $\varphi(u) \mathcal{G} \varphi(v)$. Let $\omega \geq 1$ be an integer such that $x^\omega$ is idempotent for all $x \in M$. For words $u, v \in A^*$ we set $u \to v$ if $u \equiv_\varphi v$ or if $u = pe_{m-1} \cdots e_1 z f_1 \cdots f_{m-1} q$ and $v = pe_{m-1} \cdots e_1 f_1 \cdots f_{m-1} q$ for some words $p, q, e_i, f_i, x_i, y_i, z \in A^*$ with $e_1 = 1 = f_1$ and $e_{i+1} = (e_i \cdots e_1 z f_1 \cdots f_i x_i)^\omega$ and $f_{i+1} = (y_i e_i \cdots e_1 z f_1 \cdots f_i)^\omega$. One can think of $\to$ as a semi-Thue system induced by equality in $M$ and the identity for $\mathbf{W}_{m-1}$. We let $\overset{*}{\leftrightarrow}$ be the equivalence relation generated by $\to$. That is $u \overset{*}{\leftrightarrow} v$ if there exists $w_0, \ldots, w_k$ with $u = w_0$, $v = w_k$ and with $w_i \to w_{i+1}$ or $w_{i+1} \to w_i$ for all $0 \leq i < k$. The relation $\overset{*}{\leftrightarrow}$ is a congruence on $A^*$ with finite index, and $A^*/\overset{*}{\leftrightarrow}$ is in $\mathbf{W}_{m-1}$ by definition. Note that $\varphi(u) = \varphi(v)$ implies $u \overset{*}{\leftrightarrow} v$ and therefore, $A^*/\overset{*}{\leftrightarrow}$ is a quotient of $M$. In particular, we have $u^{2\omega} \overset{*}{\leftrightarrow} u^\omega$. By assumption there exists $n$ such that $u \equiv_{m-1,n} v$ implies $u \overset{*}{\leftrightarrow} v$.

Suppose $u \equiv_{m-1,n} v$, $x \mathcal{R} xu$, and $y \mathcal{L} vy$. By choice of $n$ there exists $u = w_0, \ldots, w_k = v$ with $w_i \to w_{i+1}$ or $w_{i+1} \to w_i$. We claim that if $t \to w$ and $x \mathcal{R} xt$ and $y \mathcal{L} ty$, then $xty \equiv_\varphi xwy$. This is trivial if $t \equiv_\varphi w$. Suppose $t = pt'q$ and $w = pw'q$ with $t' = e_{m-1} \cdots e_1 z f_1 \cdots f_{m-1}$ and $w' = e_{m-1} \cdots e_1 f_1 \cdots f_{m-1}$ with $e_1 = 1 = f_1$ and $e_{i+1} = (e_i \cdots e_1 z f_1 \cdots f_i x_i)^\omega$ and $f_{i+1} = (y_i e_i \cdots e_1 z f_1 \cdots f_i)^\omega$. We have $xp \mathcal{R} xpt'$ because $x \mathcal{R} xt$. Hence there exists $x_{m-1} \in A^*$ such that $xp \equiv_\varphi xpt' x_{m-1} \equiv_\varphi upe_m$ with $e_m = (t' x_{m-1})^\omega = (e_{m-1} \cdots e_1 z f_1 \cdots f_{m-1} x_{m-1})^\omega$. Symmetrically, for some $y_{m-1} \in A^*$ and $f_m = (y_{m-1} e_{m-1} \cdots e_1 z f_1 \cdots f_{m-1})^\omega$ we get $qy \equiv_\varphi f_m qy$. With $M \in \mathbf{W}_m$ we conclude

$$\begin{aligned} xty =& \ xpe_{m-1} \cdots e_1 z f_1 \cdots f_{m-1} qy \\ \equiv_\varphi& \ xpe_m e_{m-1} \cdots e_1 z f_1 \cdots f_{m-1} f_m qy \\ \equiv_\varphi& \ xpe_m e_{m-1} \cdots e_1 \ f_1 \cdots f_{m-1} f_m qy \\ \equiv_\varphi& \ xpe_{m-1} \cdots e_1 f_1 \cdots f_{m-1} qy = xwy. \end{aligned}$$

If $t \to w$, then either $t \equiv_\varphi w$ or $\alpha(t) = \alpha(w)$. Therefore by Lemma 1, we have $x \mathcal{R} xt$ if and only if $x \mathcal{R} xw$, and $y \mathcal{L} ty$ if and only if $y \mathcal{L} wy$ whenever $t \to w$. Thus $x \mathcal{R} xw_i$ and $y \mathcal{L} w_i y$ for all $0 \leq i \leq k$. The above claim now yields $xuy = xw_0 y \equiv_\varphi xw_1 y \equiv_\varphi \cdots \equiv_\varphi xw_k y = xvy$. □

The next lemma exploits the properties of monoids in **DA** given in Lemma 1.

**Lemma 7** *Let $\varphi : A^* \to M$ be a homomorphism with $M \in \mathbf{DA}$ and let $u \equiv_{1,2|M|-2} v$ for $u, v \in A^*$. Then there exist rankers $r \in X_A^*$ and $s \in Y_A^*$, and factorizations $u = u_0 a_1 u_1 \cdots a_\ell u_\ell$ and $v = v_0 a_1 v_1 \cdots a_\ell v_\ell$ with $a_i \in A$ and the following properties:*

1. *$|r| \leq |M| - 1$ and $|s| \leq |M| - 1$,*
2. *if $t$ is a nonempty prefix of either $r$ or $s$, then there exists $i \in \{1, \ldots, \ell\}$ such that $t(u) = |u_0 a_1 \cdots u_{i-1} a_i|$ and $t(v) = |v_0 a_1 \cdots v_{i-1} a_i|$, and for all $i \in \{1, \ldots, \ell\}$ there exists a nonempty prefix $t$ of $r$ or $s$ such that $t(u) = |u_0 a_1 \cdots u_{i-1} a_i|$ and $t(v) = |v_0 a_1 \cdots v_{i-1} a_i|$,*
3. *$\varphi(u_0) \mathcal{R} 1$ and $\varphi(u_0 a_1 \cdots u_{i-1} a_i u_i) \mathcal{R} \varphi(u_0 a_1 \cdots u_{i-1} a_i)$ for all $i \in \{1, \ldots, \ell\}$,*
4. *$\varphi(v_\ell) \mathcal{L} 1$ and $\varphi(v_{i-1} a_i v_i \cdots a_\ell v_\ell) \mathcal{L} \varphi(a_i v_i \cdots a_\ell v_\ell)$ for all $i \in \{1, \ldots, \ell\}$.*

**Remark 1** *We note the following observations regarding Lemma 7: In the premise, we have $u \equiv_{1,2|M|-2} v$ if and only if $u \rhd_{1,2|M|-2} v$ if and only if $u \lhd_{1,2|M|-2} v$. We have $\ell \leq 2|M|-2$. The $a_i$ are exactly the positions visited by $r$ and $s$ on both words $u$ and $v$, and their relative order (with respect to the rankers $r$ and $s$) is the same. The factorization of $u$ is a refinement of its $\mathcal{R}$-factorization, and the factorization of $v$ is a refinement of its $\mathcal{L}$-factorization.* ◇

*Proof (Lemma 7):* We extend relations $\mathcal{G}$ on $M$ to words $u, v \in A^*$ by setting $u \mathcal{G} v$ if $\varphi(u) \mathcal{G} \varphi(v)$. Consider the $\mathcal{R}$-factorization of $u$, i.e., let $u = x_0 b_1 \cdots x_{k-1} b_k x_k$ with $b_i \in A$ be such that $1 \mathcal{R} x_0$ and

$$x_0 b_1 \cdots x_{i-1} >_\mathcal{R} x_0 b_1 \cdots x_{i-1} b_i \mathcal{R} x_0 b_1 \cdots x_{i-1} b_i x_i$$

for all $1 \leq i \leq k$. Here we let $x >_\mathcal{R} y$ in $M$ if $x \geq_\mathcal{R} y$ but not $x \mathcal{R} y$. We have $k \leq |M|-1$ since the number of $\mathcal{R}$-classes is bounded by $|M|$. The positions of the $b_i$'s in $u$ are *red*. Symmetrically, consider the $\mathcal{L}$-factorization of $v$, i.e., let $v = y'_0 c_1 \cdots y'_{k'-1} c_{k'} y'_{k'}$ with $c_i \in A$ be such that $1 \mathcal{L} y'_{k'}$ and

$$y'_j \cdots c_{k'} y'_{k'} >_\mathcal{L} c_j y'_j \cdots c_{k'} y'_{k'} \mathcal{L} y'_{j-1} c_j y'_j \cdots c_{k'} y'_{k'}$$

for all $1 \leq j \leq k'$. As before we have $k' \leq |M|-1$. The positions of the $c_j$'s in $v$ are *blue*. The next step is to transfer the red positions from $u$ to $v$ and the blue positions from $v$ to $u$ such that the relative order of the colored positions are the same on $u$ and $v$. By Lemma 1 we have $b_i \notin \alpha(x_{i-1})$ and hence $|x_0 b_1 \cdots x_{i-1} b_i| = \mathsf{X}_{b_1} \cdots \mathsf{X}_{b_i}(u)$, i.e., the positions of $b_i$ in $u$ are exactly the ones visited by the ranker $r = \mathsf{X}_{b_1} \cdots \mathsf{X}_{b_k} \in X_A^*$. By assumption, $r$ is also defined on $v$ and we let $v = y_0 b_1 \cdots y_{k-1} b_k v_k$ such that $|v_0 b_1 \cdots v_{i-1} b_i| = \mathsf{X}_{b_1} \cdots \mathsf{X}_{b_i}(v)$ for all $i$. In $v$ the positions of these $b_i$ are red. Similarly, we transfer the blue positions to $u$. We have $c_j \notin \alpha(y'_j)$ by Lemma 1 and thus $|y'_0 c_1 \cdots v'_{j-1} c_j| = \mathsf{Y}_{c_{k'}} \cdots \mathsf{Y}_{c_j}(v)$. Let $s = \mathsf{Y}_{c_{k'}} \cdots \mathsf{Y}_{c_1}$. Let $u = x'_0 c_1 \cdots x'_{k'-1} c_{k'} x'_{k'}$ with $|x'_0 c_1 \cdots x'_{j-1} c_j| = \mathsf{Y}_{c_{k'}} \cdots \mathsf{Y}_{c_j}(u)$. The positions of these $c_j$ are blue.

Colored positions of $u$ and $v$ are either red or blue or both. The relative order between positions with the same color is trivially the same on $u$ as on $v$. Since all red positions are reachable by a ranker in $X_A^*$ of depth at most $|M|-1$ and all blue positions are reachable by a ranker in $Y_A^*$ of depth at most $|M|-1$, Lemma 3 shows that the relative order between all colored positions is the same on $u$ as on $v$. Therefore it is possible to factorize $u = u_0 a_1 u_1 \cdots a_\ell u_\ell$ and $v = v_0 a_1 v_1 \cdots a_\ell v_\ell$ with $a_i \in A$ such that these factorizations satisfy property (2). Moreover, since these factorizations are refinements of the original $\mathcal{R}$-/$\mathcal{L}$-factorizations, they also satisfy properties (3) and (4). □

**Lemma 8** *Let $m \geq 3$, let $n \in \mathbb{N}$, and let $\varphi : A^* \to M$ be a homomorphism with $M \in \mathbf{DA}$. Suppose for all $u, v, x, y \in A^*$ the following implication holds:*

$$u \equiv_{m-1,n} v, \quad \varphi(x) \mathcal{R} \varphi(xu), \quad \varphi(y) \mathcal{L} \varphi(vy) \quad \Rightarrow \quad xuy \equiv_\varphi xvy.$$

*Then $u \equiv_{m,n+2|M|-1} v$ implies $u \equiv_\varphi v$ for all $u, v \in A^*$.*

*Proof:* Let $u = u_0 a_1 u_1 \cdots a_\ell u_\ell$ and $v = v_0 a_1 v_1 \cdots a_\ell v_\ell$ be the factorizations given by Lemma 7, and let them be defined by the rankers $r \in X_A^*$ and $s \in Y_A^*$ with $|r|+|s| \leq 2|M|-2$. Lemma 5 shows $u_i \equiv_{m-1,n} v_i$ for all $i$. Repeated application of the assumption in order to translate $v$ into $u$ yields

$$\begin{aligned} v = &\ v_0 a_1 v_1 a_2 \cdots v_{\ell-1} a_\ell v_\ell \\ \equiv_\varphi &\ u_0 a_1 v_1 a_2 \cdots v_{\ell-1} a_\ell v_\ell \\ \equiv_\varphi &\ u_0 a_1 u_1 a_2 \cdots v_{\ell-1} a_\ell v_\ell \\ &\ \vdots \\ \equiv_\varphi &\ u_0 a_1 u_1 a_2 \cdots u_{\ell-1} a_\ell v_\ell \\ \equiv_\varphi &\ u_0 a_1 u_1 a_2 \cdots u_{\ell-1} a_\ell u_\ell = u. \end{aligned}$$

Note that the substitution rules $v_i \mapsto u_i$ are $\varphi$-invariant only if applied from left to right. □

We are now ready to prove the inclusion $\mathbf{R}_m \vee \mathbf{L}_m \subseteq \mathbf{W}_m$.

**Theorem 1** *For $m \geq 2$ we have $\mathbf{R}_m \vee \mathbf{L}_m = \mathbf{W}_m$.*

*Proof:* We have $\mathbf{R}_m \subseteq \mathbf{W}_m$ and $\mathbf{L}_m \subseteq \mathbf{W}_m$ by Proposition 1. Since $\mathbf{W}_m$ is a variety, we see that $\mathbf{R}_m \vee \mathbf{L}_m \subseteq \mathbf{W}_m$. For the converse inclusion we prove the following claim: For every homomorphism $\varphi : A^* \to M$ with $M \in \mathbf{W}_m$ there exists $n \in \mathbb{N}$ such that $u \equiv_{m,n} v$ implies $u \equiv_\varphi v$. The inclusion $\mathbf{W}_m \subseteq \mathbf{R}_m \vee \mathbf{L}_m$ then follows from this claim by the first statement in Corollary 1.

The proof of the claim is by induction on $m$. Almeida and Azevedo have shown $\mathbf{R}_2 \vee \mathbf{L}_2 = \mathbf{W}_2$, see [2, 1]. Thus, by the second statement in Corollary 1, the claim holds for $m = 2$. Let now $m > 2$ and let $\varphi : A^* \to M$ be a homomorphism with $M \in \mathbf{W}_m$. By induction, for every homomorphism $\mu : A^* \to N$ with $N \in \mathbf{W}_{m-1}$ there exists $n' \in \mathbb{N}$ such that $u \equiv_{m-1,n'} v$ implies $u \equiv_\mu v$. By Lemma 6, there exists $n \in \mathbb{N}$ such that for all $u, v, x, y \in A^*$ the following implication holds:

$$u \equiv_{m-1,n} v, \ \varphi(x) \mathcal{R} \varphi(xu), \ \varphi(y) \mathcal{L} \varphi(vy) \ \Rightarrow \ xuy \equiv_\varphi xvy.$$

With Lemma 2 and Lemma 8 we see that $u \equiv_{m,n+2|M|-1} v$ implies $u \equiv_\varphi v$. This proves the claim. □

Since membership in $\mathbf{W}_m$ is decidable, Theorem 1 immediately yields the following decidability result.

**Corollary 2** *For every given integer $m \geq 2$ and every given finite monoid $M$ it is decidable whether $M$ is in $\mathbf{R}_m \vee \mathbf{L}_m$.* □

## 6 Separating the Join Levels from the Intersection Levels

For every $m \geq 2$ we show that there is a language which is recognized by a monoid in $\mathbf{R}_{m+1} \cap \mathbf{L}_{m+1}$, but not by a monoid in $\mathbf{R}_m \vee \mathbf{L}_m$. This last statement relies on Theorem 1; and the membership in $\mathbf{R}_{m+1} \cap \mathbf{L}_{m+1}$ uses condensed rankers and Proposition 2.

**Theorem 2** *For all $m \geq 2$ we have $\mathbf{R}_m \vee \mathbf{L}_m \subsetneq \mathbf{R}_{m+1} \cap \mathbf{L}_{m+1}$.*

*Proof:* We have $\mathbf{R}_m \cup \mathbf{L}_m \subseteq \mathbf{R}_{m+1} \cap \mathbf{L}_{m+1}$, see [12, Corollary 3.19]. Since $\mathbf{R}_{m+1} \cap \mathbf{L}_{m+1}$ is a variety, we obtain $\mathbf{R}_m \vee \mathbf{L}_m \subseteq \mathbf{R}_{m+1} \cap \mathbf{L}_{m+1}$. For showing $\mathbf{R}_m \vee \mathbf{L}_m \neq \mathbf{R}_{m+1} \cap \mathbf{L}_{m+1}$ we give a language $L_{m,m}$ which is recognized by a monoid in $\mathbf{R}_{m+1} \cap \mathbf{L}_{m+1}$ but not by a monoid in $\mathbf{R}_m \vee \mathbf{L}_m$. Let $b_2, b_3, \ldots, c_2, c_3, \ldots, d$ be distinct letters. We set $A_i = \{b_2, \ldots, b_{i-1}, c_2, \ldots, c_{i-1}, d\}$, $B_i = A_i \cup \{b_i\}$ and $C_i = A_i \cup \{c_i\}$. Consider the languages $L_{k,\ell}$ given by

$$L_{k,\ell} = B_k^* b_k \cdots B_2^* b_2 \, d \, c_2 C_2^* \cdots c_\ell C_\ell^*.$$

Note that all languages $L_{k,\ell}$ are unambiguous, *i.e.*, every word $w \in L_{k,\ell}$ has a unique factorization $w = u_k b_k \cdots u_2 b_2 \, d \, c_2 v_2 \cdots c_\ell v_\ell$ with $u_i \in B_i^*$ and $v_j \in C_j^*$. To this end, observe that in every word in $L_{k,\ell}$, the marker $b_k$ is the last occurrence of this letter and symmetrically, the marker $c_\ell$ is the first occurrence.

We claim that for all $m \geq 1$ and every alphabet $A$ with $C_{m+1} \subseteq A$ there exist finite sets of rankers $S, T \subseteq R_{m+1} \cap X_A Z_A^*$ such that

$$L_{m,m+1} = \bigcap_{r \in S} L_c(r) \setminus \bigcup_{r \in T} L_c(r).$$

Then, by symmetry, for each $m \geq 1$ and every alphabet $A$ with $B_{m+1} \subseteq A$ there exist sets of rankers $S, T \subseteq R_{m+1} \cap Y_A Z_A^*$ such that we have $L_{m+1,m} = \bigcap_{r \in S} L_c(r) \setminus \bigcup_{r \in T} L_c(r)$. For $m = 1$ we have $L_{1,2} = d \, c_2 \{c_2, d\}^*$ and we may choose

$$S = \{\mathsf{X}_{c_2} \mathsf{Y}_d\},$$
$$T = \{\mathsf{X}_{c_2} \mathsf{Y}_d \mathsf{Y}_d\} \cup \{\mathsf{X}_a \mid a \in A \setminus C_2\}.$$

Let now $m \geq 2$. We have

$$L_{m,m+1} = L_{m,m-1} c_m C_m^* c_{m+1} C_{m+1}^*.$$

By induction there exist sets $S', T' \subseteq R_m \cap Y_A Z_A^*$ with $L_{m,m-1} = \bigcap_{r \in S'} L_c(r) \setminus \bigcup_{r \in T'} L_c(r)$. We set

$$S = \mathsf{X}_{c_m} S' \cup \{\mathsf{X}_{c_m} \mathsf{X}_{c_{m+1}}\},$$
$$T = \mathsf{X}_{c_m} T' \cup \{\mathsf{X}_a \mid a \in A \setminus C_{m+1}\} \cup$$
$$\{\mathsf{X}_{c_m} \mathsf{X}_{c_{m+1}} \mathsf{Y}_a \mid a \in A \setminus C_m\}.$$

Here, $rS$ for a ranker $r$ and a set of rankers $S$ is the set of rankers $\{rs \mid s \in S\}$. Note that $c_m$ does not appear in $L_{m,m-1} \subseteq B_m^*$ and that all rankers in $S'$ and $T'$ start with a Y-modality. Therefore, the rankers from $\mathsf{X}_{c_m} S'$ and $\mathsf{X}_{c_m} T'$ in the above definition "relativize" the ranker-description of $L_{m,m-1}$ to the factor before the first $c_m$-position, *i.e.*, if $r$ is a ranker and $u, v \in A^*$ are such that $c_m \notin \alpha(u)$, then $u \in L_c(\mathsf{Y}_a r)$ if and only if $uc_m v \in L_c(\mathsf{X}_{c_m} \mathsf{Y}_a r)$.

Consider now the language $L_{m,m} = B_m^* b_m L_{m-1,m} \subseteq A^*$ for some alphabet $A$. By the above claim, we get finite sets $S', T'$ of rankers in $R_m \cap X_A Z_A^*$ such that $L_{m-1,m} = \bigcap_{r \in S'} L_c(r) \setminus \bigcup_{r \in T'} L_c(r)$. Then setting $S = \mathsf{Y}_{b_m} S'$ and $T = \{\mathsf{Y}_{b_m} \mathsf{Y}_a \mid a \notin B_m\} \cup \mathsf{Y}_{b_m} T'$ yields $L_{m,m} = \bigcap_{r \in S} L_c(r) \setminus \bigcup_{r \in T} L_c(r)$. This shows that there exists $n \in \mathbb{N}$ such that $L_{m,m}$ is a union of $\triangleleft_{m+1,n}$-classes, *i.e.*, $L_{m,m}$ is recognized by $A^*/\triangleleft_{m+1,n}$. Proposition 2 shows that $L_{m,m}$ is recognized by a monoid in $\mathbf{L}_{m+1}$. By symmetry, $L_{m,m}$ is also recognized by a monoid in $\mathbf{R}_{m+1}$. It follows that the syntactic monoid of $L_{m,m}$ is in $\mathbf{R}_{m+1} \cap \mathbf{L}_{m+1}$. In particular, $L_{m,m}$ is recognized by a monoid in $\mathbf{R}_{m+1} \cap \mathbf{L}_{m+1}$.

Next, we show that there is no monoid in $\mathbf{R}_m \vee \mathbf{L}_m$ recognizing $L_{m,m}$. Consider a homomorphism $\varphi : A^* \to M$ with $M \in \mathbf{W}_m$ and choose $n$ such that all $n$th powers are idempotent in $M$. Let $e_1 = f_1 = 1$ and for $i \geq 2$ let $e_i = (e_{i-1} \cdots e_1 d f_1 \cdots f_{i-1} b_i)^n$, and let $f_i = (c_i e_{i-1} \cdots e_1 d f_1 \cdots f_{i-1})^n$. Then $e_i \in B_i^* b_i$ and $f_i \in c_i C_i^*$. Hence $u = e_m \cdots e_1 d f_1 \cdots f_m \in L_{m,m}$. Using unambiguity of $L_{m,m}$, an elementary verification shows that $v = e_m \cdots e_1 f_1 \cdots f_m \notin L_{m,m}$. Since $\varphi(u) = \varphi(v)$ by $M \in \mathbf{W}_m$, the homomorphism $\varphi$ does not recognize the language $L_{m,m}$. By Theorem 1 we see that $M \notin \mathbf{R}_m \vee \mathbf{L}_m$ for every monoid $M$ recognizing $L_{m,m}$. □

## 7 Applications

In this section, we relate the join levels of the Trotter-Weil hierarchy with two other hierarchies. First, we consider the hierarchy of deterministic and codeterministic products, starting with languages of the form $B^*$ for $B \subseteq A$; and we show that this hierarchy is decidable. This result essentially follows from Theorem 1 and from Pin's characterization of deterministic and codeterministic products [14]. Our second application is unambiguous interval temporal logic (unambiguous ITL). It has been introduced by Lodaya, Pandya, and Shah [13] as an expressively complete logic for **DA**. We show that the direction alternation depth within unambiguous ITL is decidable. This result heavily relies on combinatorial properties of rankers given by Kufleitner and Weil [12].

### 7.1 Deterministic and Codeterministic Products

A language of the form $L = L_0 a_1 L_1 \cdots a_k L_k$ with $L_i \in \mathcal{V}$ is called a *monomial* over a class of languages $\mathcal{V}$. The monomial $L$ is *unambiguous* if every word $u \in L$ has a unique factorization $u = u_0 a_1 u_1 \cdots a_k u_k$ such that $u_i \in L_i$; it is *deterministic* if for every word $u \in L$ and every $i \in \{1, \ldots, k\}$ there is a unique prefix of $u$ which is in $L_0 a_1 \cdots L_{i-1} a_i$; and it is *codeterministic* (also called *reverse deterministic*) if for every word $u \in L$ and every $i \in \{1, \ldots, k\}$ there is a unique suffix of $u$ in $a_i L_i \cdots a_k L_k$. Every deterministic or codeterministic monomial is unambiguous.

If $\mathcal{V}$ is a class of languages, then the *deterministic closure* $\mathcal{V}^{det}$ of $\mathcal{V}$ (resp. the *codeterministic closure* $\mathcal{V}^{codet}$ of $\mathcal{V}$) is the Boolean closure of the deterministic (resp. codeterministic) monomials over $\mathcal{V}$. Alternating between closure under deterministic and codeterministic monomials and between closure under Boolean operations yields the following hierarchy: $\mathcal{W}_1$ contains all Boolean combinations of languages $B^*$ for $B \subseteq A$, and $\mathcal{W}_{m+1}$ consist of all Boolean combinations of deterministic and of codeterministic monomials over $\mathcal{W}_m$, i.e., $\mathcal{W}_{m+1}$ is the Boolean closure of $\mathcal{W}_m^{det} \cup \mathcal{W}_m^{codet}$. Since $\mathcal{V}^{det}$ and $\mathcal{V}^{codet}$ are varieties if $\mathcal{V}$ is a variety [14], each of the language classes $\mathcal{W}_m$ is a variety. The next proposition shows that $\mathcal{W}_m$ corresponds to the level $\mathbf{R}_m \vee \mathbf{L}_m$ of the Trotter-Weil hierarchy. Decidability of $\mathcal{W}_m$ then follows by Corollary 2.

**Proposition 3** *Let $L \subseteq A^*$ and let $m \geq 2$. Then $L$ is in $\mathcal{W}_m$ if and only if $L$ is recognized by a monoid in $\mathbf{R}_m \vee \mathbf{L}_m$.*

*Proof:* The proof is by induction on $m$, starting with $m = 1$. For this purpose, we set $\mathbf{R}_1 = \mathbf{L}_1 = \mathbf{J}_1$. Then $\mathbf{R}_2 = \mathbf{K} \textcircled{m} \mathbf{L}_1$ and $\mathbf{L}_2 = \mathbf{D} \textcircled{m} \mathbf{R}_1$, see *e.g.* [15]. An easy observation is that $\mathcal{W}_1$ corresponds to $\mathbf{J_1}$, see *e.g.* [4]. Thus the claim holds for $m = 1$. Let now $m > 1$. By induction, $\mathcal{W}_{m-1}$ corresponds to $\mathbf{R}_{m-1} \vee \mathbf{L}_{m-1}$. By Pin's characterization of deterministic and codeterministic products [14], $\mathcal{W}_{m-1}^{det}$ corresponds to $\mathbf{K} \circledm (\mathbf{R}_{m-1} \vee \mathbf{L}_{m-1})$. Similarly, $\mathcal{W}_{m-1}^{codet}$ corresponds to $\mathbf{D} \circledm (\mathbf{R}_{m-1} \vee \mathbf{L}_{m-1})$. Since $\mathbf{R}_{m-1} \vee \mathbf{L}_{m-1} \subseteq \mathbf{R}_m \cap \mathbf{L}_m$, we obtain

$$\mathbf{R}_m \subseteq \mathbf{K} \circledm (\mathbf{R}_{m-1} \vee \mathbf{L}_{m-1}) \subseteq \mathbf{K} \circledm \mathbf{R}_m = \mathbf{R}_m,$$
$$\mathbf{L}_m \subseteq \mathbf{D} \circledm (\mathbf{R}_{m-1} \vee \mathbf{L}_{m-1}) \subseteq \mathbf{D} \circledm \mathbf{L}_m = \mathbf{L}_m.$$

Since the Boolean closure $\mathcal{W}_m$ of $\mathcal{W}_{m-1}^{det} \cup \mathcal{W}_{m-1}^{codet}$ is the smallest variety containing both $\mathcal{W}_{m-1}^{det}$ and $\mathcal{W}_{m-1}^{codet}$, we see that $\mathcal{W}_m$ corresponds to $\mathbf{R}_m \vee \mathbf{L}_m$. □

### 7.2 Alternation within Unambiguous ITL

The syntax of *unambiguous interval temporal logic* (unambiguous ITL) is as follows. Formulae are built from the atoms $\top$ for *true* and $\bot$ for *false*, and if $\varphi$ and $\psi$ are formulae in unambiguous ITL, then so are

$$\neg \varphi \mid \varphi \vee \psi \mid \varphi \wedge \psi \mid \varphi \, \mathsf{F}_a \, \psi \mid \varphi \, \mathsf{L}_a \, \psi$$

for every letter $a \in A$. The modality $\mathsf{F}_a$ stands for "First $a$-position" and $\mathsf{L}_a$ is for "Last $a$-position". Usually, models are finite words and an interval of positions. For the purpose of this paper, we use only word models (without intervals). The semantics is as follows. Every word $u \in A^*$ models $\top$, written as $u \models \top$, and no word models $\bot$. Boolean combinations are as usual. The semantics of $\mathsf{F}_a$ and $\mathsf{L}_a$ is given by

$$u \models \varphi \, \mathsf{F}_a \, \psi \;\Leftrightarrow\; a \in \alpha(u) \text{ and for } u = u_0 a u_1 \text{ with } a \notin \alpha(u_0)$$
$$\text{we have } u_0 \models \varphi \text{ and } u_1 \models \psi,$$
$$u \models \varphi \, \mathsf{L}_a \, \psi \;\Leftrightarrow\; a \in \alpha(u) \text{ and for } u = u_0 a u_1 \text{ with } a \notin \alpha(u_1)$$
$$\text{we have } u_0 \models \varphi \text{ and } u_1 \models \psi.$$

That is, for the formula $\varphi \, \mathsf{F}_a \, \psi$, the model is "split" at the first $a$-position and $\varphi$ and $\psi$ are interpreted over the resulting left and right factors, respectively. The modality $\mathsf{L}_a$ is left-right dual. For a formula $\varphi$ we set $L(\varphi) = \{u \in A^* \mid u \models \varphi\}$. Then

$$L(\varphi \, \mathsf{F}_a \, \psi) = (L(\varphi) \cap B^*) a L(\psi), \qquad L(\varphi \, \mathsf{L}_a \, \psi) = L(\varphi) a (L(\psi) \cap B^*)$$

where $B = A \setminus \{a\}$. We introduce the parameter $t$ ("*turns*") of a formula: we let $t(\top) = t(\bot) = 0$, $t(\neg \varphi) = t(\varphi)$ and $t(\varphi \vee \psi) = t(\varphi \wedge \psi) = \max\{t(\varphi), t(\psi)\}$; and for the temporal modalities we let

$$t(\varphi \, \mathsf{F}_a \, \psi) = \max\{t(\varphi)+1, t(\psi)\}, \qquad t(\varphi \, \mathsf{L}_a \, \psi) = \max\{t(\varphi), t(\psi)+1\}.$$

The parameter $t$ defines the number of direction alternations in a formula (more precisely, the number of blocks of directions). We shall also need the following parameter $d(\varphi)$ capturing the nesting depth of $\mathsf{F}_a$ and $\mathsf{L}_a$ of the formula: let $d(\top) = d(\bot) = 0$, $d(\neg \varphi) = d(\varphi)$ and $d(\varphi \vee \psi) = d(\varphi \wedge \psi) = \max\{d(\varphi), d(\psi)\}$; and for the temporal modalities we set

$$d(\varphi \, \mathsf{F}_a \, \psi) = \max\{d(\varphi)+1, d(\psi)+1\}, \quad d(\varphi \, \mathsf{L}_a \, \psi) = \max\{d(\varphi)+1, d(\psi)+1\}.$$

Let $\text{ITL}_{m,n}$ contain all unambiguous ITL-formulae $\varphi$ with $t(\varphi) \leq m$ and $d(\varphi) \leq n$. Let $\text{ITL}_m = \bigcup_n \text{ITL}_{m,n}$.

Next we show that agreement of words $u, v \in A^*$ on $\mathrm{ITL}_{m,n}$-formulae is the same as agreement on condensed rankers in $R_{m,n}$. We write $u \approx_{m,n} v$ if for all $\varphi \in \mathrm{ITL}_{m,n}$ we have

$$u \models \varphi \;\Leftrightarrow\; v \models \varphi.$$

For every fixed alphabet, the set $\mathrm{ITL}_{m,n}$ is finite up to equivalence. Thus $\approx_{m,n}$ is a finite index congruence.

**Proposition 4** *Let $m, n \in \mathbb{N}$ and $u, v \in A^*$. Then $u \approx_{m,n} v$ if and only if $u \equiv_{m,n} v$.*

*Proof:* The proof is by induction on $m + n$. If $m = 0$ or $n = 0$, then the claim is vacuously true. In the sequel we assume $m, n \geq 1$. First suppose $u \approx_{m,n} v$. By symmetry, it suffices to show $u \triangleright_{m,n} v$. We use the following characterization of $u \triangleright_{m,n} v$, see [12, Proposition 3.6]. For $m, n \geq 1$ we have $u \triangleright_{m,n} v$ if and only if:

- $\alpha(u) = \alpha(v)$ and $u \triangleleft_{m-1,n-1} v$, and
- for all $a \in \alpha(u)$, the factorizations $u = u_0 a u_1$ and $v = v_0 a v_1$ with $a \notin \alpha(u_0) \cup \alpha(v_0)$ satisfy $u_0 \triangleleft_{m-1,n-1} v_0$ and $u_1 \triangleright_{m,n-1} v_1$.

Now, $\alpha(u) = \alpha(v)$ since $a \in \alpha(u)$ is equivalent to $u \models \top \mathsf{F}_a \top$. We have $u \triangleleft_{m-1,n-1} v$ by induction. Let $a \in \alpha(u)$, and let $u = u_0 a u_1$ and $v = v_0 a v_1$ with $a \notin \alpha(u_0) \cup \alpha(v_0)$. For all formulae $\varphi \in \mathrm{ITL}_{m-1,n-1}$ we have $u_0 \models \varphi$ if and only if $u \models \varphi \mathsf{F}_a \top$ if and only if $v \models \varphi \mathsf{F}_a \top$ if and only if $v_0 \models \varphi$. Thus $u_0 \approx_{m-1,n-1} v_0$. Similarly, we see that $u_1 \approx_{m,n-1} v_1$. By induction, we obtain $u_0 \triangleleft_{m-1,n-1} v_0$ and $u_1 \triangleright_{m,n-1} v_1$. Therefore, we have $u \triangleright_{m,n} v$.

Next, suppose $u \equiv_{m,n} v$. We claim that for all $\varphi \in \mathrm{ITL}_{m,n}$ we have $u \models \varphi$ if and only if $v \models \varphi$. The proof is by induction on the formula. For atoms the claim is true, and for Boolean combinations, the claim follows by induction. Let $u = u_0 a u_1$ and $v = v_0 a v_1$ with $a \notin \alpha(u_0) \cup \alpha(v_0)$. For all $m \geq 2$ we have $u_0 \equiv_{m-1,n-1} v_0$ and $u_1 \equiv_{m,n-1} v_1$ by [12, Proposition 3.6 and Lemma 3.7]. Moreover, for $m = 1$ we have $u_1 \triangleright_{1,n-1} v_1$, and this is equivalent to $u_1 \triangleleft_{1,n-1} v_1$, see *e.g.* [12, Proposition 4.1]. Since the assertion $u_0 \equiv_{0,n-1} v_0$ is always true, we have $u_0 \equiv_{m-1,n-1} v_0$ and $u_1 \equiv_{m,n-1} v_1$ for all $m \geq 1$. Now, induction yields $u_0 \approx_{m-1,n-1} v_0$ and $u_1 \approx_{m,n-1} v_1$. Therefore if $\varphi \mathsf{F}_a \psi \in \mathrm{ITL}_{m,n}$, then $u \models \varphi \mathsf{F}_a \psi$ if and only if both $u_0 \models \varphi$ and $u_1 \models \psi$ if and only if both $v_0 \models \varphi$ and $v_1 \models \psi$ if and only if $v \models \varphi \mathsf{F}_a \psi$. Note that $\varphi \in \mathrm{ITL}_{m-1,n-1}$ and $\psi \in \mathrm{ITL}_{m,n-1}$. The case of $\varphi \mathsf{L}_a \psi \in \mathrm{ITL}_{m,n}$ is symmetric. □

**Corollary 3** *Let $L \subseteq A^*$ and $m \geq 2$. Then $L$ is definable in $\mathrm{ITL}_m$ if and only if $L$ is recognized by a monoid in $\mathbf{R}_m \vee \mathbf{L}_m$. In particular, it is decidable whether a given regular language is definable in $\mathrm{ITL}_m$.*

*Proof:* If $L$ is definable in $\mathrm{ITL}_m$, then there exists $n \in \mathbb{N}$ such that $L$ is a union of $\approx_{m,n}$-classes, and by Proposition 4 it is a union of $\equiv_{m,n}$-classes. Thus Corollary 1 shows that $L$ is recognized by $A^*/\equiv_{m,n} \in \mathbf{R}_m \vee \mathbf{L}_m$.

Suppose now $L$ is recognized by $\varphi : A^* \to M$ onto $M \in \mathbf{R}_m \vee \mathbf{L}_m$. Corollary 1 implies that there exists $n \in \mathbb{N}$ such that $M$ is a quotient of $A^*/\equiv_{m,n}$. Consequently, by Proposition 4, the language $L$ is a union of $\approx_{m,n}$-classes. The claim follows since each such class is a Boolean combination of $\mathrm{ITL}_{m,n}$-formulae. Note that $\mathrm{ITL}_{m,n}$ is closed under Boolean operations.

The decidability result follows immediately by Corollary 2. □